\documentclass[aps,superscriptaddress,showpacs,nofootinbib]{revtex4-1}

\usepackage{amscd}
\usepackage{amsmath,amssymb,amsthm,mathrsfs,amsfonts,dsfont}
\usepackage{subfigure, epsfig}
\usepackage{braket}
\usepackage{bm}
\usepackage{enumerate}

\usepackage{graphicx}
\usepackage{epsfig}
\usepackage{subfigure}
\usepackage{color}
\usepackage[10pt]{moresize}

\usepackage[bookmarks = false]{hyperref}
\usepackage{bm}
\usepackage[all]{hypcap}
\usepackage{colortbl}
\usepackage{booktabs}
\usepackage{enumitem}
\usepackage{mathrsfs}
\usepackage{multirow}
\usepackage{upgreek}    
\usepackage{rotating}
\usepackage{setspace}
\usepackage[resetlabels]{multibib}
\newcites{sec}{Reference}

\newcommand\oprod[2]{\ensuremath{|#1\rangle\langle#2|}}
\newcommand\mean[1]{\ensuremath{\langle #1 \rangle}}
\newcommand{\msection}[1]{\vspace{\baselineskip}{\centering \textbf{#1}\\}\vspace{0.5\baselineskip}}

\begin{document}

\title{Field Test of Twin-Field Quantum Key Distribution through Sending-or-Not-Sending over 428~km}

\author{Hui Liu}
\affiliation{Hefei National Laboratory for Physical Sciences at Microscale and Department of Modern Physics, University of Science and Technology ofChina, Hefei, Anhui 230026, People’s Republic of China}
\affiliation{CAS Center for Excellence in Quantum Information and Quantum Physics, University of Science and Technology of China, Hefei, Anhui 230026, People’s Republic of China}

\author{Cong Jiang}
\affiliation{Jinan Institute of Quantum Technology, Jinan, Shandong 250101, People’s Republic of China}

\author{Hao-Tao Zhu}
\author{Mi Zou}
\affiliation{Hefei National Laboratory for Physical Sciences at Microscale and Department of Modern Physics, University of Science and Technology ofChina, Hefei, Anhui 230026, People’s Republic of China}
\affiliation{CAS Center for Excellence in Quantum Information and Quantum Physics, University of Science and Technology of China, Hefei, Anhui 230026, People’s Republic of China}

\author{Zong-Wen Yu}
\affiliation{State Key Laboratory of Low Dimensional Quantum Physics, Department of Physics, Tsinghua University, Beijing 100084, People’s Republic of China}
\affiliation{Data Communication Science and Technology Research Institute, Beijing 100191, People’s Republic of China}

\author{Xiao-Long Hu}
\author{Hai Xu}
\affiliation{State Key Laboratory of Low Dimensional Quantum Physics, Department of Physics, Tsinghua University, Beijing 100084, People’s Republic of China}

\author{Shizhao Ma}
\author{Zhiyong Han}
\affiliation{Jinan Institute of Quantum Technology, Jinan, Shandong 250101, People’s Republic of China}

\author{Jiu-Peng Chen}
\affiliation{Hefei National Laboratory for Physical Sciences at Microscale and Department of Modern Physics, University of Science and Technology ofChina, Hefei, Anhui 230026, People’s Republic of China}
\affiliation{CAS Center for Excellence in Quantum Information and Quantum Physics, University of Science and Technology of China, Hefei, Anhui 230026, People’s Republic of China}

\author{Yunqi Dai}
\author{Shi-Biao Tang}
\affiliation{QuantumCTek Corporation Limited, Hefei, Anhui 230088, People’s Republic of China}

\author{Weijun Zhang}
\author{Hao Li}
\author{Lixing You}
\author{Zhen Wang}
\affiliation{State Key Laboratory of Functional Materials for Informatics, Shanghai Institute of Microsystem and Information Technology, Chinese Academy of Sciences, Shanghai 200050, People’s Republic of China}

\author{Fei Zhou}
\affiliation{Jinan Institute of Quantum Technology, Jinan, Shandong 250101, People’s Republic of China}

\author{Qiang Zhang}
\affiliation{Hefei National Laboratory for Physical Sciences at Microscale and Department of Modern Physics, University of Science and Technology ofChina, Hefei, Anhui 230026, People’s Republic of China}
\affiliation{CAS Center for Excellence in Quantum Information and Quantum Physics, University of Science and Technology of China, Hefei, Anhui 230026, People’s Republic of China}
\affiliation{Jinan Institute of Quantum Technology, Jinan, Shandong 250101, People’s Republic of China}

\author{Xiang-Bin Wang}
\affiliation{CAS Center for Excellence in Quantum Information and Quantum Physics, University of Science and Technology of China, Hefei, Anhui 230026, People’s Republic of China}
\affiliation{Jinan Institute of Quantum Technology, Jinan, Shandong 250101, People’s Republic of China}
\affiliation{State Key Laboratory of Low Dimensional Quantum Physics, Department of Physics, Tsinghua University, Beijing 100084, People’s Republic of China}

\author{Teng-Yun Chen}
\author{Jian-Wei Pan}
\affiliation{Hefei National Laboratory for Physical Sciences at Microscale and Department of Modern Physics, University of Science and Technology ofChina, Hefei, Anhui 230026, People’s Republic of China}
\affiliation{CAS Center for Excellence in Quantum Information and Quantum Physics, University of Science and Technology of China, Hefei, Anhui 230026, People’s Republic of China}

\begin{abstract}
{Quantum key distribution endows people with information-theoretical security in communications. Twin-field quantum key distribution (TF-QKD) has attracted considerable attention because of its outstanding key rates over long distances. Recently, several demonstrations of TF-QKD have been realized. Nevertheless, those experiments are implemented in the laboratory, remaining a critical question about whether the TF-QKD is feasible in real-world circumstances. Here, by adopting the sending-or-not-sending twin-field QKD (SNS-TF-QKD) with the method of actively odd parity pairing (AOPP), we demonstrate a field-test QKD over 428~km deployed commercial fiber and two users are physically separated by about 300~km in a straight line. To this end, we explicitly measure the relevant properties of the deployed fiber and develop a carefully designed system with high stability. The secure key rate we achieved breaks the absolute key rate limit of repeater-less QKD. The result provides a new distance record for the field test of both TF-QKD and all types of fiber-based QKD systems. Our work bridges the gap of QKD between laboratory demonstrations and practical applications, and paves the way for intercity QKD network with high-speed and measurement-device-independent security.}
\end{abstract}

\maketitle

{\it Introduction.---}
Since Bennet and Brassard proposed the BB84 protocol~\cite{bennett1984quantum}, quantum key distribution (QKD) has been studied extensively~\cite{gisin2002quantum,scarani2009security,liao2017satellite,xu2020secure} towards its final goal of application in the real-life world. Given the fact that quantum signals cannot be amplified, the secure distance is severely limited by the channel loss. For example, considering the possible photon-number-splitting (PNS) attack, the key rate of a BB84 protocol with the imperfect single-photon source  is propositional to $\eta^2$, given the channel transmittance $\eta$.
 So far, many efforts have been made towards the more loss-tolerant QKD in practice. There are two mile-stone signs of progress towards this goal. First, the decoy-state method~\cite{H03,wang05,LMC05} can improve the key rate of coherent-state based QKD from quadratic scaling $\eta^2$ to linear scaling $\eta$, as what behaves of a perfect single-photon source. Importantly, the method can be applied to the measurement-device-independent QKD (MDI-QKD) successfully~\cite{braunstein2012side,lo2012measurement,WangPRA2016,MDI404km}. Second, the secure key rate can be further improved to the scale of the square root of the channel transmittance by using the ideal of twin-field QKD (TF-QKD)~\cite{nature18}. This method has the potential to break the known distance records of existing protocols in practical QKD and break the theoretical key rate limit of a trusted-relay-less QKD protocol known as the Pirandola-Laurenza-Ottaviani-Bianchi (PLOB) bound~\cite{PLOB2017}. 

The real-world QKD aims to physically separate users on Earth. However, despite tremendous efforts were made into fiber-based QKD field test ~\cite{peev2009the, stucki2011longterm, chen2009field, dynes2019cambridge, chen2010metropolitan,wang2010field, tang2016measurement}, the maximum fiber distance is about 130~km~\cite{chen2010metropolitan} to date. The maximal physical separation achieved between two users is about 100~km~\cite{chen2010metropolitan}, and challenges for longer distances remain. 

It is worth noting that experimental TF-QKD ~\cite{minder2019experimental, liuyang2019, wang2019beating, zhong2019proof, fang2019surpassing, chen2020sending} has advanced significantly up to a distance of more than 500~km~\cite {fang2019surpassing, chen2020sending}. However, all the experiments are implemented in the laboratory with either the simulated channel loss or the optical fiber spool, leaving a huge gap between laboratory demonstrations and practical applications. Field trial of TF-QKD remains experimentally challenging.

In this work, for the first time, we present a field-test of high-rate TF-QKD on the deployed commercial fiber (428~km length with 79.1~dB channel loss, buried underground). Furthermore, it is the longest fiber-based QKD field test without relying on trusted relays. Two users, Alice and Bob, realize the longest physical separation distance (about 300~km) in the terrestrial QKD so far, to the best of our knowledge. The secure key rate of our work breaks the absolute key rate limit of trusted-relay-less QKD. The result lays the foundation for a high-speed intercity-scale QKD network in the absence of the quantum repeater.

We adopt the sending-or-not-sending (SNS) protocol~\cite {wang2018sns} of TF-QKD with finite-key effects~\cite {jiang2019}. Besides, we apply the efficient error rejection method, known as the actively odd parity pairing (AOPP)~\cite{xu2019general} with the finite-key effects studied in Ref. \cite {jiang2019}. Given such an asymmetric channel, we adopt the asymmetric protocol \cite {huxl} to improve the secure key rate further.


{\it Protocol.---}Consider the SNS-TF-QKD protocol proposed in Ref.~\cite{wang2018sns}. Here, we implement an asymmetric 3-intensity method for decoy-state analysis. To improve the key rate, we take bit error rejection by AOPP \cite{xu2019general} in the post data processing stage. In this way, the sending probability in signal windows can be far improved and hence the number of effective events is raised greatly. As a result, the final key rate is improved a lot especially in the case of small data size with finite key effects being considered. We use the zigzag approach to take the finite-key effects in calculating the final key rate~\cite{jiang2019}.

In the protocol, Alice (Bob) randomly chooses the decoy window and signal window with probabilities $1-p_{A2}(1-p_{B2})$ and $p_{A2}(p_{B2})$, respectively. In the decoy window, both Alice and Bob prepare and send decoy pulses. In our 3-intensity protocol, there are $2$ types of decoy states in decoy windows for each party of Alice and Bob, one vacuum and one non-vacuum coherent states, of intensity $\mu_{A1}$ for Alice and $\mu_{B1}$ for Bob. Private random phase shifts of $\theta_{A}$ and $\theta_B$ are applied to each pulse. And in the signal window, Alice (Bob) decides to send out a phase-randomized weak coherent state pulse with intensity $\mu_{A2}$ ($\mu_{B2}$) or a vacuum pulse with probabilities $\epsilon_A$ $(\epsilon_B)$ and $1-\epsilon_A$ $(1-\epsilon_B)$, respectively. A $Z$ window event is defined as an event that both Alice and Bob choose the signal windows. A $Z$ window event is regarded as being effective if Charlie announces that only one detector clicked. An $X$ window event is defined as an event that both Alice’s WCS pulse is $\mu_{A1}$ and the intensity of Bob’s WCS pulse is $\mu_{B1}$ and their phases satisfy an extra phase-slice condition to reduce the observed error rate~\cite{huxl}. As shown in Ref.~\cite{huxl}, we set the condition of 
\begin{equation}
\frac{\mu_{A_1}}{\mu_{B1}}=\frac{\epsilon_A(1-\epsilon_B)\mu_{A2}e^{-\mu_{A2}}}{\epsilon_B(1-\epsilon_A)\mu_{B2}e^{-\mu_{B2}}}
\end{equation}
for the security of our asymmetric protocol.

An error in the $X$ window is defined as an effective event in the $X$ window when Charlie announces a click of right (left) while the phase difference between the pulse pair from Alice and Bob would provably cause a left (right) clicking at Charlie's measurement set-up. At a signal window, Alice (Bob) puts down a bit value 1(0) when she (he) decides sending, Alice (Bob) puts down a bit value 0(1) when she (he) decides not-sending. The values of $e_1^{ph}$ and $n_1$, the phase-flip error rate and the number of effective single-photon events in the $Z$-basis, can be calculated by the conventional decoy-state method~\cite{wang2018sns,yu2018sending}. Then we can calculate the secure key rate by the zigzag approach proposed in Ref.~\cite{jiang2019}. Calculation details are shown in the Supplemental Materials.

\begin{figure}[!h]
  \includegraphics[width=0.9\textwidth]{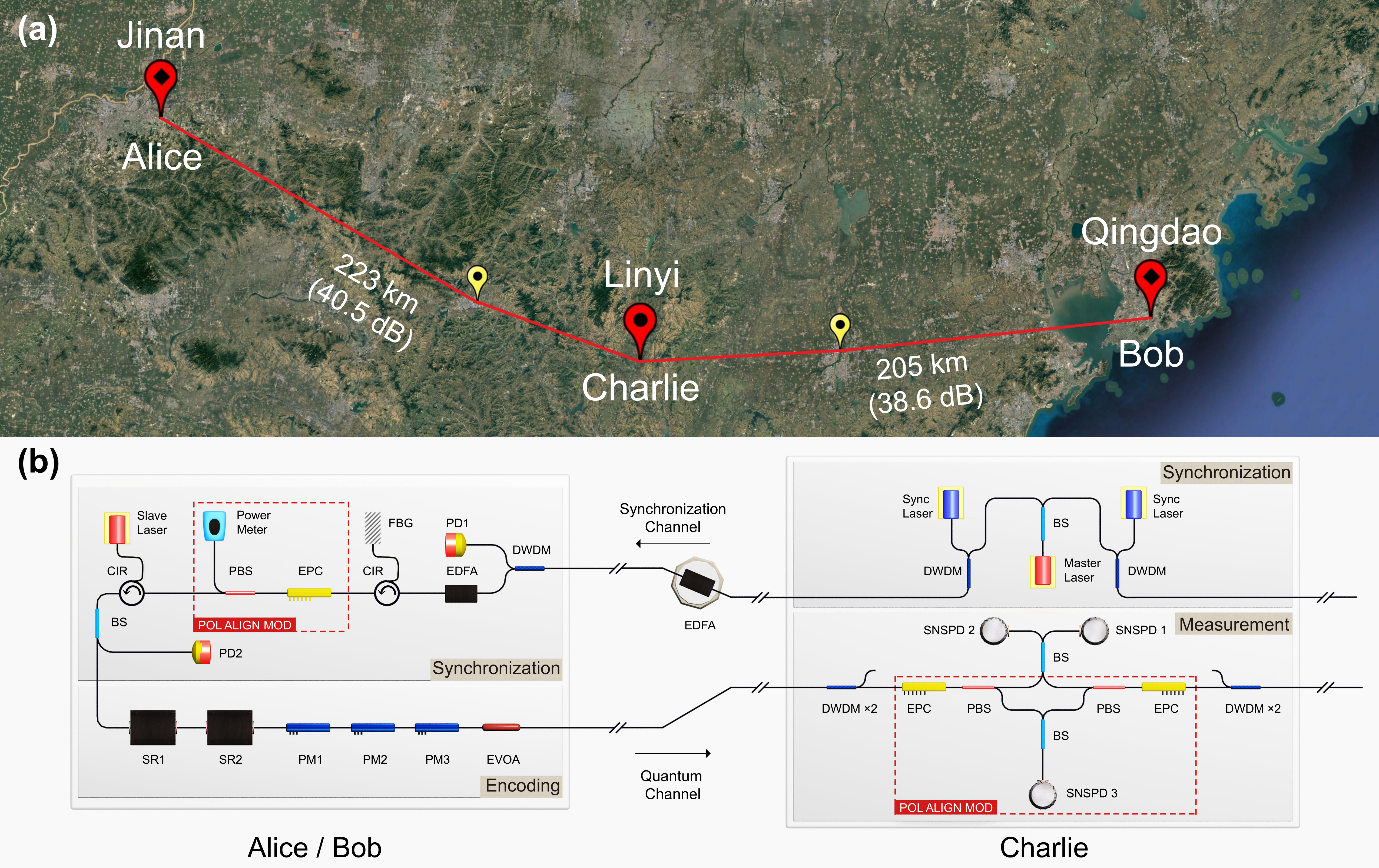}\\
  \caption{(a)	Bird's-eye view of our field test. Alice is located at the Jinan Institute of Quantum Technology (JIQT) in Jinan city (36°41'0.60" N, 117°8'10.93" E), while Bob is located at an internet data center (IDC) room in Qingdao city (36°7'24.29" N, 120°27'11.88" E). The third-party measurement is done by Charlie in a room at Linyi city (36°1'39.84" N, 118°44'50.58" E), which is 223~km from Alice and 205~km from Bob. Two yellow marks show the locations of two machine rooms at Yiyuan city (36$^\circ$ 1'12.60" N, 118$^\circ$ 2'24.16" E) and Zhucheng city (36$^\circ$ 11'59.31" N, 119$^\circ$ 4'43.58" E), respectively. An erbium-doped fiber amplifier (EDFA) is placed in each machine room to amplify the light for the clock and wavelength synchronization, of which the detail is shown in Section Experiment and the Supplemental Materials. Map data from Google, Landsat/Copernicus. (b) Illustration of the experimental set-up. A continuous-wave (cw) bright beam from a 1550.12~nm master laser is multiplexed with the pulses from two 1570~nm auxiliary synchronization lasers (Sync Lasers) in Charlie and is transmitted along the synchronization channel. At each sides of Alice and Bob, the slave laser is seeded by the cw bright beam and generates pulses with a width of 320~ps and a repetition rate of 312.5~MHz. The optical launch power of the slave laser is monitored in real-time by a watchdog photoelectric detector PD2 of Alice (Bob). Then these pulses are sent to two sagnac rings SR1-2, which is randomly prepared in one of the four intensity: strong $\mu_{r}$, high $\mu_{A2}$ ($\mu_{B2}$), moderate $\mu_{A1}$ ($\mu_{B1}$), and vacuum state. Three phase modulators PM1-3 are utilized for active phase randomization. The pulses are transmitted along the quantum channel and interfere in Charlie. The outcome of the interference is detected by the superconducting nanowire single photon detectors SNSPD1 and SNSPD2. The schematic of the polarization auto-alignment module is shown inside the red dashed rectangle. EVOA:  electrical variable optical attenuator, FBG: fiber Bragg grating, CIR: circulator, EDFA: erbium-doped fiber amplifier, EPC: electric polarization controller, DWDM: dense wavelength division multiplexer, PBS: polarizing beam splitter.}
  \label{fig:MapSetup}
\end{figure}

{\it Experiment.---}
In our field test, Alice and Bob are located in Jinan city and Qingdao city, respectively. The central relay Charlie is placed in Linyi city, as shown in FIG.~\ref{fig:MapSetup} (a). The distance between Charlie and Alice (Bob) is 223~km with 40.5~dB channel loss (205~km with 38.6~dB channel loss).

The experimental setup is comprised of the synchronization system and the encoding and measurement system, as shown in FIG.~\ref{fig:MapSetup} (b). Alice and Bob are connected by two parallel field-deployed commercial fibers (in the same optical cable) with 428~km length each, which are named synchronization channel and quantum channel respectively in the following. 

The synchronization system includes two functions: 1) the clock synchronization, of which the details are shown in the Supplemental Materials; 2) the wavelength synchronization. The first issue that makes implementation difficult is avoiding the rapid relative phase drift caused by Alice's and Bob's lasers' wavelength difference. We realize the wavelength synchronization with the assistance of the laser injection technique. A laser with 1550.12~nm wavelength and 3~kHz linewidth is placed in Charlie as the master laser. The continuous-wave (cw) bright beam is produced and injected into Alice's and Bob's slave laser. To guarantee 0~dBm cw bright beam injected into slave laser, we add four erbium-doped fiber amplifiers (EDFAs), two of which are placed in Yiyuan city (36$^\circ$1'12.60" N, 118$^\circ$2'24.16" E) and Zhucheng city (36$^\circ$11'59.31" N, 119$^\circ$4'43.58" E), respectively (as shown in FIG.~\ref{fig:MapSetup}). And the rest two are added in Alice's and Bob's apparatus. A 10~GHz fiber Bragg grating (FBG) is inserted in Alice's (Bob's) apparatus to filter the amplified spontaneous emission (ASE) noise of the EDFAs. To gain stable and high-efficiency injection, a polarization auto-alignment module is inserted before the injection. 

The pulses produced from the slave laser pass through two sagnac rings (SRs) and three phase modulators (PMs) for encoding and phase randomization in the encoding and measurement system. The pulses are attenuated to the desired levels by an electrical variable optical attenuator (EVOA) before being transmitted to Charlie through the quantum channel. In Charlie, a 50:50 BS performs a single photon interference of the incoming pulses after noise filtering. The measurement results are detected by two superconducting nanowire single photon detectors (SNSPDs) with efficiencies of 73\% and 76\%, respectively. Two polarization auto-alignment modules are employed for real-time compensate polarization drifts in the long fiber before interference. Charlie's overall detection efficiency is 28.20\%, taking into account 2.4~dB insertion loss, 30\% non-overlapping between signal pulse and detection window, and 94\% polarization alignment efficiency. The dark count of each SNSPD is about 6~Hz, corresponding to a dark count rate of $2.0 \times 10^{-9}$/pulse.

\begin{figure}[hbtp]
	\centering
    \includegraphics[width=0.5\textwidth]{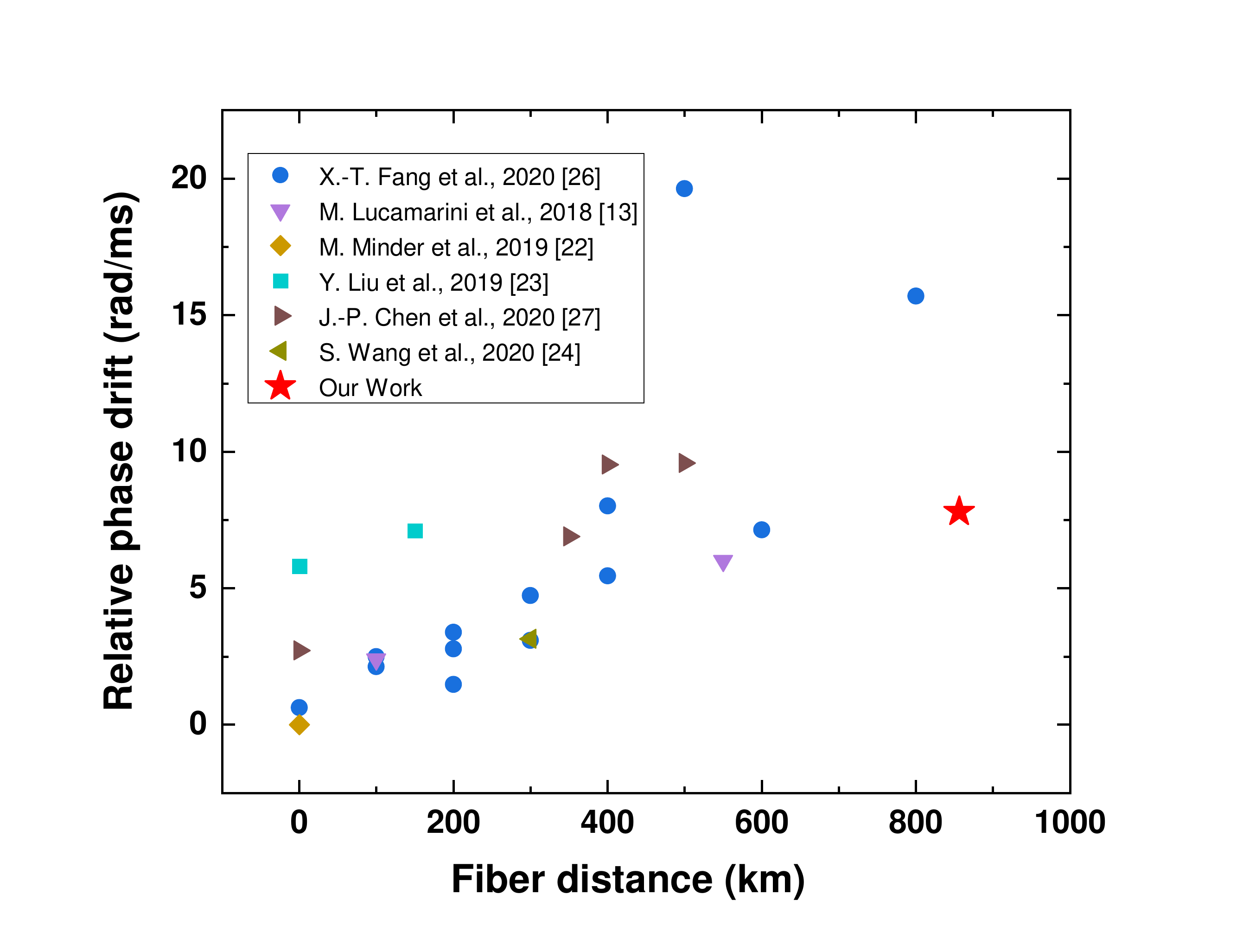}
	\caption{Relative phase drift caused by the fiber channel with different fiber distance. All results except our work are test on the optical fiber spool. In our work, the total relative phase drift is influenced by an 856 km fiber, which is 7.80 rad/ms.}
	\label{fig:RelativePhaseDrift}
\end{figure}

Another challenge we have encountered is the significant changes to the relative phase drift stemmed from the long fiber channel. A comparison of the relative phase drift in different fiber distance in previous works and our work is shown in FIG.~\ref{fig:RelativePhaseDrift}. We stress that in our work, the signal pulses produced by the slave laser inherit the global phase of the cw bright beam, which is influenced by the 428~km synchronization channel. And the signal pulses transmit along the 428~km quantum channel before interference. So the relative phase drift is influenced by the total 856~km fiber links. Fortunately, even though the relative phase drift in our field test is influenced by the longest fiber than all the previous lab works, it is not the fastest one in all works. It makes the relative phase calculation in our field test less demanding than the lab experiment in ~\cite{fang2019surpassing} over 402~km (about 800~km fiber influencing the relative phase drift). 

We verified that we could indeed estimate and compensate the relative phase drift caused by long fiber channel. In our work, Alice and Bob sacrifice a part of signal pulses as bright reference pulses periodically and send them to Charlie for relative phase calculation and apply a post selection method to the signal pulses (the detail see ~\cite{fang2019surpassing} and the Supplemental Materials). The bright reference pulses $\mu_{r}$ are set to about 450 photons per pulse at a repetition rate of 200~MHz, which results in a 5~MHz count rate of two SNSPDs for calculation and the duration time of each calculation is 20~us. The bright reference pulses will lead to noise in the long fiber channel, which is hard to avoid in the field test and the lab test. After being filtered by four 100~GHz dense wavelength division multiplexers (DWDMs) in Charlie, the remaining noise is about $1.4\times 10^{-8}$/pulse. It is an optimal trade-off optimization, and the details are shown in the Supplemental Materials.

\begin{figure}[hbtp]
	\centering
    \includegraphics[width=0.9\textwidth]{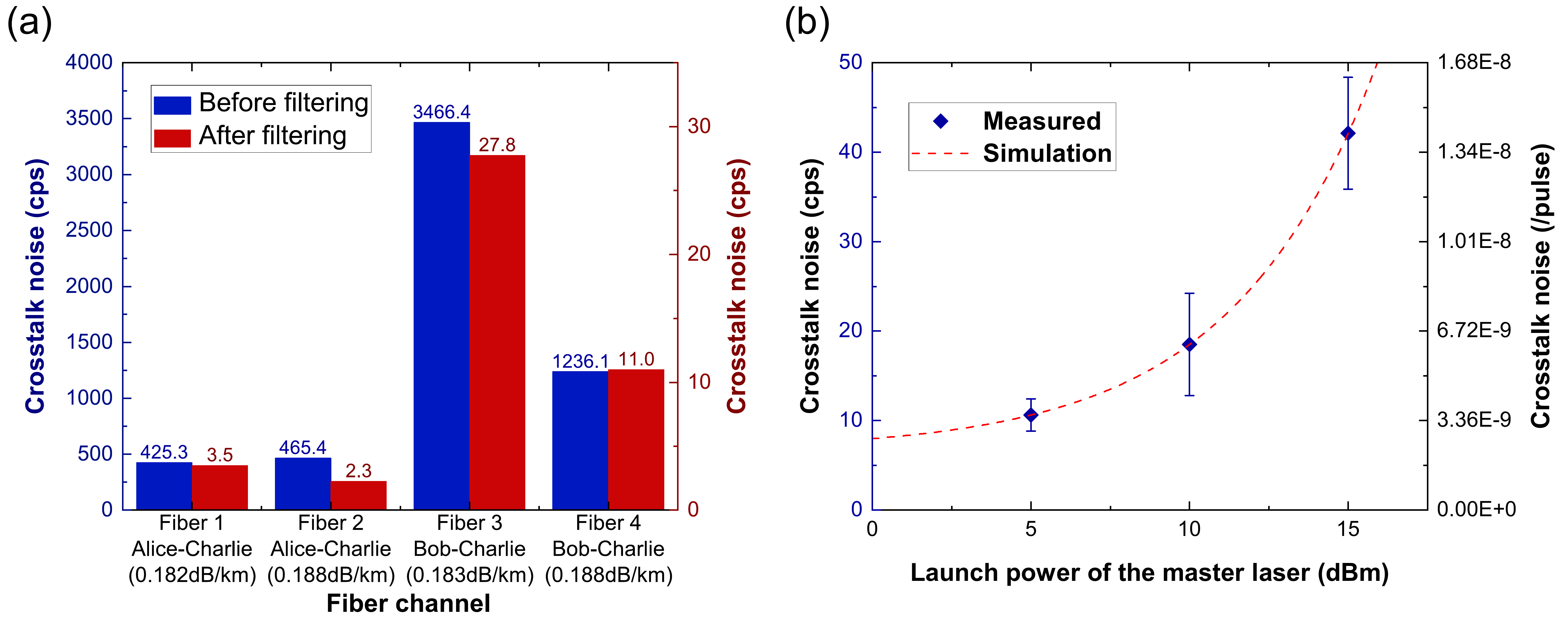}
	\caption{Characterization of the crosstalk noise. All measurements are performed under the same overall detection efficiency (28.20\%). (a) The crosstalk noise caused by the classical services running in some fibers in the optical cable. We test two available fiber channels Fiber~1 and Fiber~2 (Fiber~3 and Fiber~4) from Alice (Bob) to Charlie. The blue and red columns are the measurement results before and after filtering with two 100~GHz DWDMs, respectively. Taken the loss and the crosstalk noise of the fiber channel into account, we use Fiber~1  and Fiber~4 as the quantum channel. (b) The crosstalk noise caused by the cw bright beam in the synchronization channel with different optical launch power. Each experiment lasts 5 minutes. The experimental results are the average and variance (1 standard deviation) calculated by 144 experiments.}
	\label{fig:crosstalknoise}
\end{figure}

Besides, we face the crosstalk noise in the field test, which is never met in TF-QKD lab experiments. The quantum channel for transmitting signal pulses is in an optical cable (96 fibers included). Part of the noise proceeds from the classical services running in some fibers in the optical cable. Fortunately, it can be filtered by four DWDMs in Charlie to approximately $5.1\times 10^{-9}$/pulse, which is acceptable for us, as shown in FIG.~\ref{fig:crosstalknoise}(a). The other part of the noise is raised from the cw bright beam (same wavelength with the signal, generated from the master laser in Charlie) in the synchronization channel, which is also in the same optical cable. Thus it cannot be filtered whether spectrally or temporally. We found that the crosstalk noise becomes more ignorable as the optical launch power of the cw bright beam decrease, which is shown in FIG.~\ref{fig:crosstalknoise}(b). To suppress the noise, we reduce the optical launch power of the master laser to about 5~dBm and increase the EDFA gain appropriately, resulting in a noise level of $3.6\times 10^{-9}$/pulse. Still, a stable and high-efficiency injection can be ensured in this case.

\begin{figure}[hbtp]
	\centering
    \includegraphics[width=1\textwidth]{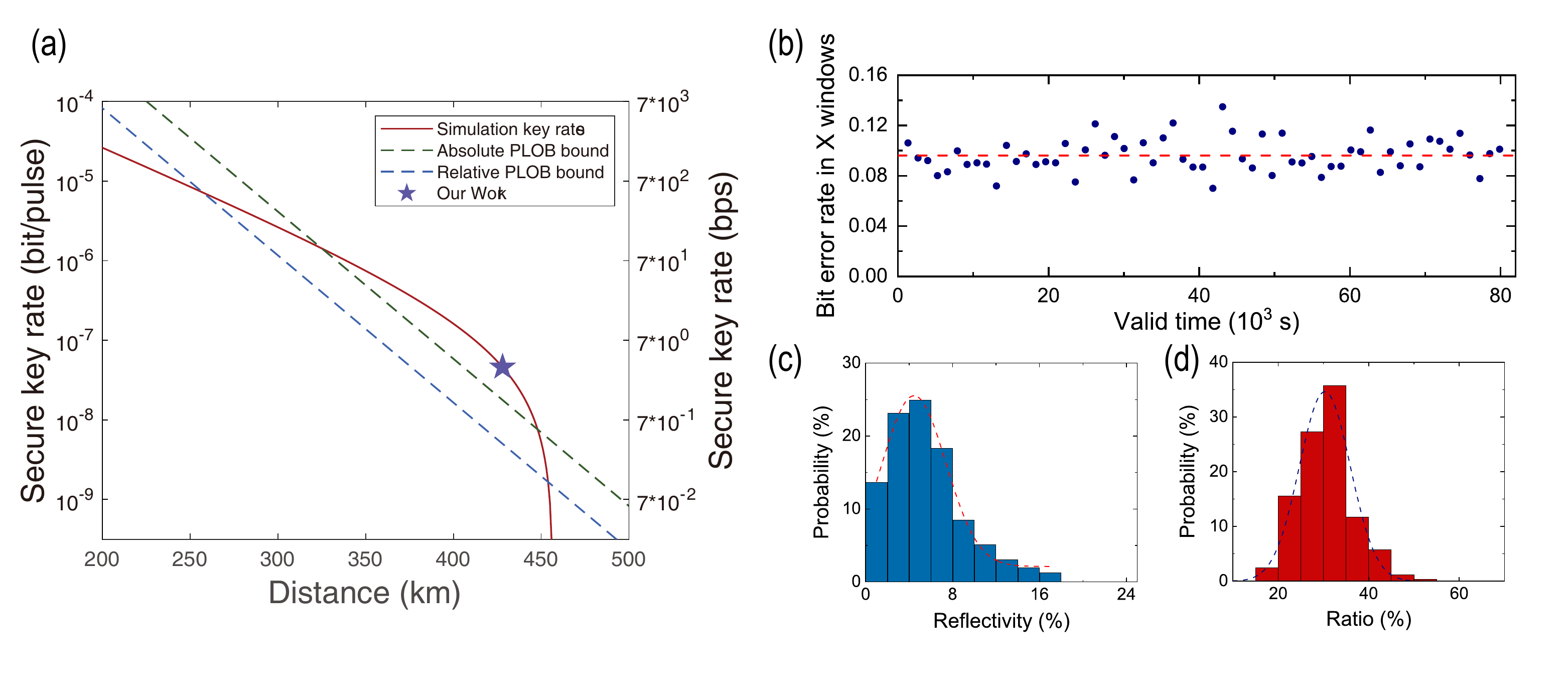}
	\caption{(a)Experimentally and simulated secure key rates. The purple pentagram point is the secure key rate of our work. The red curve is the simulation results using the parameters in Table~\ref{tab:para}. The green dashed line is the absolute PLOB bound (assuming the overall detection efficiency of Charlie $\eta_d = 1$). The blue dashed line is the relative PLOB  bound with $\eta_d = 28.20\%$. The results shows that the secure key rate obtained in our experiment is 170\% higher than the absolute PLOB bound and 859\% higher than the relative PLOB bound. (b) The bit error rate in X windows. Each data point represents the effective clikcs collected in 21.82 min on average. (c) Probability distribution of the reflectivity for the PBSs in Charlie, with real-time compensation in the overall experiment. The total efficiency of the polarization auto-alignment module is about 94\% (the detail see the Supplemental Materials). (d) Probability distribution of the ratio of the non-overlapping between signal pulse and detection window in the overall experiment. The total non-overlapping is 30\% (the detail see the Supplemental Materials). }
	\label{result}
\end{figure}

{\it Results.---}
The main system parameters are list in Table~\ref{tab:para}. In our field test, Alice and Bob sent a total of $5.59\times 10^{12}$ pulse pairs, and get $2.79\times 10^7$ sifted key bits  in the $Z$-basis, including $27.84\%$ error bits. According to the method shown in the Supplement Materials and the data acquired in the experiment, there are at least $1.29\times 10^7$ untagged bits in the sifted keys, corresponding to an $11.07\%$ phase flip error rate before AOPP. After AOPP, $5.84\times 10^6$ keys are survived in which contains 0.69\% error bits. These values are in agreement very well with the theoretically expected values. The number of untagged bits is $2.38\times 10^6$ with a corresponding phase flip error rate is $20.24\%$. With the finite-key effect being taken into consideration, we finally obtain a secure key rate of $4.80\times 10^{-8}$/pulse (corresponding to 3.36~bps), which is $170\%$ higher than the  absolute PLOB bound and $859\%$ higher than the relative PLOB bound. FIG.~\ref{result} shows the performance of our work, in terms of the simulation key rates, the achieved secure key rate and the total efficiency of the polarization auto-alignment module and arrival time synchronization.

\begin{table}[h]
\caption{List of the main experimental parameters used in the numerical simulation: total number of signal pulse pairs $N$, overall dark count probability $p_d$, overall detection efficiency of Charlie $\eta_d$,  misalignment-error probability of $X$-basis $e_d^X$, loss coefficient of the quantum channel from Alice (Bob) to Charlie $\alpha_{AC}$($\alpha_{BC}$) in dB/km, quantum channel distance from Alice (Bob) to Charlie $L_{AC} (L_{BC})$ in km, error-correction efficiency $f$ and failure probability $\epsilon$.}
\begin{spacing}{1.5}
\vspace{0.1cm}
\begin{tabular}{p{2cm}<{\centering}p{2cm}<{\centering}p{1.5cm}<{\centering}p{1.5cm}<{\centering}p{1.5cm}<{\centering}p{1.5cm}<{\centering}p{1.5cm}<{\centering}p{1.5cm}<{\centering}p{1.5cm}<{\centering}p{1.5cm}<{\centering}}
\hline\hline

 $N$ &   $p_d$ & $\eta_d$ &    $\alpha_{AC}$ &       $\alpha_{BC}$ & $L_{AC}$ & $L_{BC}$ &   $e_d^X$  &     $f$ &    $\epsilon$     \\
\hline
 $5.59\times 10^{12}$ & $2.50\times10^{-8}$ &  28.20\%  &   0.182 &     0.188 &   223 &  205 &   8\% &       1.1 &   $10^{-10}$     \\
\hline\hline
\end{tabular}
\end{spacing}
\label{tab:para}
\end{table}

{\it Conclusions.---}
Applying the SNS protocol~\cite {wang2018sns}, we have performed the first field test of high-rate TF-QKD over the deployed commercial fiber, in which 3.36~bps secure key rate was generated over 428~km. It is the longest distance of the terrestrial real-word QKD without relying on trusted relays at present and pushes the separation between two users beyond 300~km. The result demonstrated in our experiment exhibits the feasibility of the trusted-relay-less QKD in practical circumstances between cities. It motivates future demonstration of high-speed intercity-scale QKD network in the absence of the quantum repeater. Further extensions to higher key rates include increasing the system's repetition, utilizing the fiber link with lower attenuation and less noise, and enhancing the laser and detector's performance.

This work was supported by the National Key R\&D Program of China (2017YFA0303903), the Chinese Academy of Science, the National Fundamental Research Program, the National Natural Science Foundation of China (grants 11875173, 61875182 and 11674193) and Anhui Initiative in Quantum Information Technologies and Fundamental Research Funds for the Central Universities (WK2340000083).

%
\clearpage

\setcounter{figure}{0}
\setcounter{table}{0}
\setcounter{equation}{0}

\onecolumngrid

\global\long\def\theequation{S\arabic{equation}}
\global\long\def\thefigure{S\arabic{figure}}
\renewcommand{\thetable}{S\arabic{table}}
\renewcommand{\arraystretch}{0.6}

\normalsize
\msection{SUPPLEMENTARY INFORMATION}

\section{The SNS-TF-QKD protocol with odd-parity error rejection}
\subsection{The theory}
The SNS-TF-QKD protocol with asymmetric parameters is used to perform the experiment, and the theory of actively odd-parity paring (AOPP) with finite key effects is used to extract the final key.
 
In this protocol, Alice and Bob will repeat the following process $N$ times to obtain a series of data:
\noindent At each time window, Alice (Bob) randomly decides whether it is a decoy window with probability $1-p_{A2}$ ($1-p_{B2}$), or a signal window with probability $p_{A2}$ ($p_{B2}$). If it is a signal window, with probability $\epsilon_A$ ($\epsilon_B$), Alice (Bob) randomly prepares a phase-randomized weak coherent state (WCS) pulse with intensity $\mu_{A2}$ ($\mu_{B2}$), and denotes it as bit $1$ ($0$); with probability $1-\epsilon_A$ ($1-\epsilon_B$), Alice (Bob) prepares a vacuum pulse, and denotes it as bit $0$ ($1$). If it is a decoy window, Alice (Bob) randomly prepares a vacuum pulse or coherent state pulse $\ket{e^{i\theta_{A1}}\sqrt{\mu_{A1}}}$ ( $\ket{e^{i\theta_{B1}}\sqrt{\mu_{B1}}}$) with probabilities $1-p_{A1}$ and $p_{A1}$ ($1-p_{B1}$ and $p_{B1}$), respectively, where $\theta_{A1}$ and $\theta_{B1}$ are different in different windows, and are random in $[0,2\pi)$. We set the condition in Eq.~(\ref{equ:condition}) for the security.
\begin{equation}\label{equ:condition}
   \frac{\mu_{A1}}{ \mu_{B1}}= \frac{\epsilon_\text{A}(1-\epsilon_\text{B}) \mu_{A2}  e^{-\mu_{A2}}}{\epsilon_\text{B}(1-\epsilon_\text{A})  \mu_{B2} e^{-\mu_{B2}}}.
\end{equation}
Then Alice and Bob send their prepared pulses to Charlie who is assumed to perform interferometric measurements on the received pulses and then announce the results to Alice and Bob. If one and only one detector clicks in the measurement process, Charlie also tells Alice and Bob which detector it was, and Alice and Bob take it as a one-detector heralded event.

Then Alice and Bob announce the mode they used in each time window in the public channel. A time window that both Alice and Bob determined to be a signal window is called a $Z$ window. One-detector-heralded events in $Z$ windows are called effective events. Alice and Bob get two $n_t$-bit strings, $Z_A$ and $Z_B$, as formed by the corresponding bits of effective events of the $Z$ windows. We denote the bit flip error rate of strings $Z_A$ and $Z_B$ as $E$. Strings $Z_A$ and $Z_B$ will be used to extract the secure final keys. The pulse intensities in each $Z$ window, that is the decision of whether to send or not send, is kept private, but the intensities of other pulses in each window would be publicly announced after Alice and Bob finished mode calibration. For a time window that the intensity of Alice's WCS pulse is $\mu_{A1}$ and the intensity of Bob's WCS pulse is $\mu_{B1}$, Alice and Bob also announce the phase information $\theta_{A1}$ and $\theta_{B1}$ in the public channel, and if the phases of the WCS pulses satisfy
\begin{equation}
1-\vert \cos(\theta_{A1}-\theta_{B1}-\psi_{AB})\vert\le \lambda,
\end{equation}

it is called an $X$ window. Here, $\psi_{AB}$ can take an arbitrary value, which can be different from time to time as Alice and Bob like, so as to obtain a satisfactory key rate for the protocol~\cite{liu2019experimental}. $\lambda$ is a positive value close to $0$, and would be optimized to obtain the highest key rate. One-detector-heralded events in $X$ windows are called effective events.

The AOPP method is used to reduce the errors of raw key strings $Z_A$ and $Z_B$ before the error correction and privacy amplification processes. In AOPP, Bob first actively pairs the bits $0$ with bits $1$ of the raw key string $Z_B$, and get $n_g$ pairs. And then Alice computes the parities of those $n_g$ pairs and announces them to Bob, then they keep the pairs with parity value $1$ at both sides and discard the pairs with parity $0$ at Alice’s side.
Finally, Alice and Bob randomly keep one bit from each survived pair and form two new $n_t^\prime$-bits strings, which would be used to extract the secure final keys. The formula of the key rate $R$ of AOPP is 
\begin{equation}\label{r2}
\begin{split}
R=&\frac{1}{N}\{n_1^\prime[1-h(e_{1}^{\prime ph})]-fn_t^\prime h(E^\prime)-2\log_2{\frac{2}{\varepsilon_{cor}}}\\
&-2\log_2{\frac{1}{\sqrt{2}\varepsilon_{PA}\hat{\varepsilon}}}\}.
\end{split}
\end{equation}
where $n_1^\prime$ is the number of the untagged bits after AOPP, $e_{1}^{\prime ph}$ is the phase flip error rate of untagged bits after AOPP, $h(x)=-x\log_2x-(1-x)\log_2(1-x)$ is the Shannon entropy, $E^\prime$ is the bit-flip error rate of the remaining bits after AOPP, $\varepsilon_{cor}$ is the failure probability of error correction, $\varepsilon_{PA}$ is the failure probability of privacy amplification, and $\hat{\varepsilon}$ is the coefficient while using the chain rules of smooth min- and max- entropy~\cite{vitanov2013chain}.

$n_1^\prime$ and $e_{1}^{\prime ph}$ are values after AOPP. As shown in Ref.~\cite{jiang2020zigzag}, we can calculate $n_1^\prime$ and $e_{1}^{\prime ph}$ by taking the number of untagged bits and phase flip error rate before AOPP as input values. The calculation details are shown in the next section.

\subsection{The calculation method}
To clearly show the calculation method, we denote the vacuum source, the WCS source with intensity $\mu_{A1}$, the WCS source with intensity $\mu_{A2}$ of Alice by $o,x$ and $y$. Similarly, we denote the vacuum source, the WCS source with intensity $\mu_{B1}$, the WCS source with intensity $\mu_{B2}$ of Bob by $o^\prime,x^\prime$, and $y^\prime$. We denote the number of pulse pairs of source $\kappa\zeta(\kappa=o,x,y;\zeta=o^\prime,x^\prime,y^\prime)$ sent out in the whole protocol by $N_{\kappa\zeta}$, and the total number of one-detector heralded events of source $\kappa\zeta$ by $n_{\kappa\zeta}$. We define the counting rate of source $\kappa\zeta$ by $S_{\kappa\zeta}=n_{\kappa\zeta}/N_{\kappa\zeta}$, and the corresponding expected value by $\mean{S_{\kappa\zeta}}$. With all those definitions, we have
\begin{equation}
\begin{split}
N_{oo^\prime}=&(1-p_{A2})(1-p_{B2})(1-p_{A1})(1-p_{B1})N\\
N_{ox^\prime}=&(1-p_{A2})(1-p_{B2})(1-p_{A1})p_{B1}N\\
N_{xo^\prime}=&(1-p_{A2})(1-p_{B2})p_{A1}(1-p_{B1})N\\
N_{oy^\prime}=&(1-p_{A2})p_{B2}(1-p_{A1})\varepsilon_{B}N\\
N_{yo^\prime}=&p_{A2}(1-p_{B2})\varepsilon_A(1-p_{B1})N
\end{split}
\end{equation}

As sources $x,y,x^\prime,y^\prime$ are phase-randomized WCS sources, they are actually the classical mixture of different photon number states~\cite{hu2019general}. Thus we can use the decoy-state method to calculate the lower bounds of the expected values of the counting rate of states $\oprod{01}{01}$ and $\oprod{10}{10}$, which are
\begin{align}
\label{s01mean}\mean{\underline{s_{01}}}&= \frac{\mu_{B2}^{2}e^{\mu_{B1}}\mean{S_{ox^\prime}}-\mu_{B1}^{2}e^{\mu_{B2}}\mean{S_{oy^\prime}}-(\mu_{B2}^{2}-\mu_{B1}^{2})\mean{S_{oo^\prime}}}{\mu_{B2}\mu_{B1}(\mu_{B2}-\mu_{B1})},\\
\mean{\underline{s_{10}}}&= \frac{\mu_{A2}^{2}e^{\mu_{A1}}\mean{S_{xo^\prime}}-\mu_{A1}^{2}e^{\mu_{A2}}\mean{S_{yo^\prime}}-(\mu_{A2}^{2}-\mu_{A1}^{2})\mean{S_{oo^\prime}}}{\mu_{A2}\mu_{A1}(\mu_{A2}-\mu_{A1})}.
\end{align}
Then we can get the lower bound of the expected value of the counting rate of untagged photons
\begin{equation}
\mean{\underline{s_1}}=\frac{\mu_{A1}}{\mu_{A1}+\mu_{B1}}\mean{\underline{s_{10}}}+\frac{\mu_{B1}}{\mu_{A1}+\mu_{B1}}\mean{\underline{s_{01}}},
\end{equation}
and the lower bound of the expected value of the untagged bits $1$, $\mean{\underline{n_{10}}}$, and untagged bits $0$, $\mean{\underline{n_{01}}}$
\begin{align}
\mean{\underline{n_{10}}}=Np_{A2}p_{B2}\epsilon_A(1-\epsilon_B)\mu_{A2}e^{-\mu_{A2}}\mean{\underline{s_{10}}},\\
\mean{\underline{n_{01}}}=Np_{A2}p_{B2}\epsilon_B(1-\epsilon_A)\mu_{B2}e^{-\mu_{B2}}\mean{\underline{s_{10}}}.
\end{align}

The error counting rate of the $X$ windows $T_{X}$, can be used to estimate the upper bound of the expected value of the phase flip error rate $\mean{\overline{e_1^{ph}}}$. The criterion of error events in $X$ windows are shown in Ref.~\cite{hu2019general}. We denote the number of total pulses with intensities $\mu_{A1}$ and $\mu_{B1}$ sent out in the $X$ windows by $N_{X}$, and the number of corresponding error events by $m_{X}$, then we have
\begin{equation}
T_{X}=\frac{m_{X}}{N_{X}}.
\end{equation}
Then we have
\begin{equation}\label{e1}
\mean{\overline{e_1^{ph}}}=\frac{\mean{T_{X}}-e^{-\mu_{A1}-\mu_{B1}}\mean{S_{oo^\prime}}/2}{e^{-\mu_{A1}-\mu_{B1}}(\mu_{A1}+\mu_{B1})\mean{\underline{s_1}}},
\end{equation}
where $\mean{T_{X}}$ is the expected value of $T_{X}$. 

By taking the estimated values before AOPP, $\mean{\underline{n_{10}}},\mean{\underline{n_{01}}}$ and $\mean{\overline{e_1^{ph}}}$ as input values, we can calculate $n_1^\prime$ and $e_{1}^{\prime ph}$ by the method proposed in Ref.~\cite{jiang2020zigzag}. We have the related formulas as follows.
\begin{subequations}
\begin{align}
&u=\frac{n_g}{2n_{odd}},\quad \underline{n_{1}}=\varphi^L(\mean{\underline{n_{10}}}+\mean{\underline{n_{01}}}),\\
&\underline{n_{10}}=\varphi^L(\mean{\underline{n_{10}}}),\underline{n_{01}}=\varphi^L(\mean{\underline{n_{01}}}),\\
&n=\varphi^L\left(\frac{{\underline{n_{1}}}}{n_t}\frac{{\underline{n_{1}}}}{n_t}\frac{un_t}{2}\right),\\
&k=u\underline{n_{1}}-2n,\\
&r=\frac{2n+k}{k}\ln\frac{3k^2}{\varepsilon(r,k)},\\
&\bar{M}=\varphi^U(2n\mean{\overline{e_1^{ph}}}),\\
&\mean{e_{\tau}}=\frac{\bar{M}}{2n-r}\\
&\bar{M}_s=\varphi^U[(n-r)\mean{e_{\tau}}(1-\mean{e_{\tau}})]+r,
\end{align}
\end{subequations}
where ${n_{odd}}$ is the number of pairs with odd-parity if Bob randomly groups all the bits in $Z_B$ two by two, $n_{odd}$ is observed values, $\varepsilon(r,k)=10^{-10}$ is the trace distance while using the exponential de Finetti’s representation theorem, and $\varphi^U(x),\varphi^L(x)$ are the upper and lower bounds while using Chernoff bound~\cite{chernoff1952measure} to estimate the real values according to the expected values.

Finally, we can calculate $n_1^\prime$ and $e_{1}^{\prime ph}$ by
\begin{align}
&n_1^\prime=\varphi^L\left(\frac{{\underline{n_{01}}}}{n_{t0}}\frac{{\underline{n_{10}}}}{n_{t1}}n_g\right),\\
&e_{1}^{\prime ph}=\frac{2\bar{M}_s}{n_1^\prime},
\end{align}   
where $n_{t0}$ is the number of all bit $0 $ and $n_{t1}$ is the number of all bit $1$ in string $Z_B$.

\section{DETAILS OF THE EXPERIMENT}
\subsection{The relative phase calculation}
Instead of actively phase-locking, we calculate the relative phase straightforwardly and apply a post-selection method for effective events when Alice and Bob both choose X-window. The total relative phase is influenced by an 856~km fiber, which is 7.80~rad/ms in our work. 

\begin{figure*}[hbtp]
	\centering
    \includegraphics[width=0.9\textwidth]{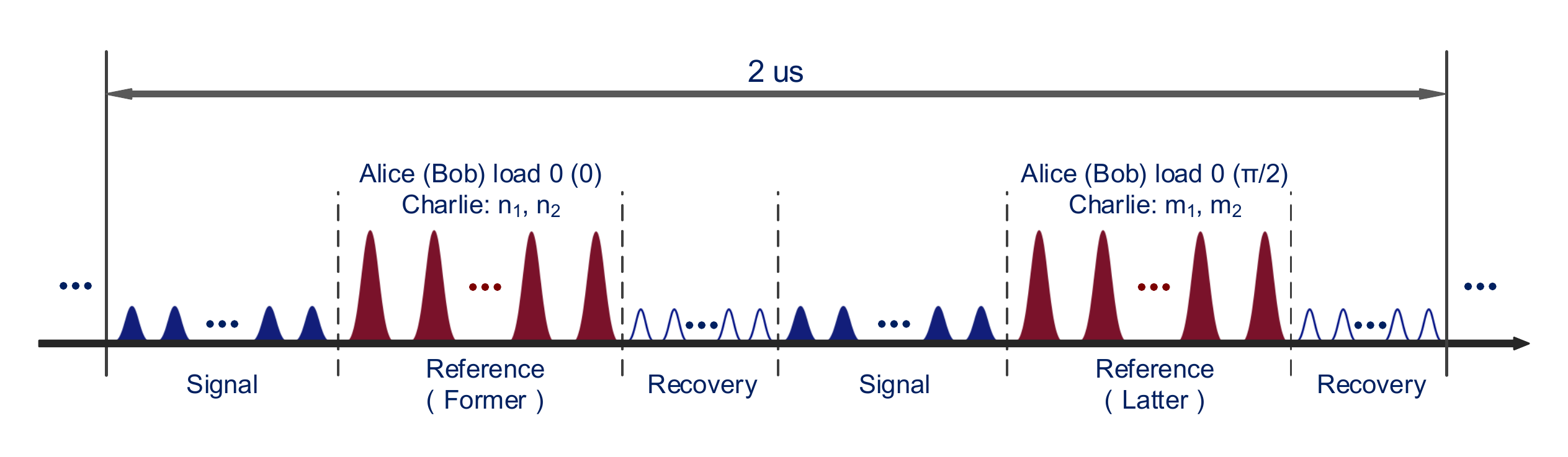}
	\caption{Time sequence in one cycle. Alice and Bob send the signal pulses in the signal region. The bright reference pulses are sent in the Reference region. Alice (Bob) loads 0 (0) and 0 ($\pi/2$) phase in the former and latter reference region, respectively. The vacuum state are send in the recovery region followed by each reference region to avoid influencing signal detection in Charlie. After interference, Charlie record the counts of SNSPD1 and SNSPD2 in the former (latter) reference region, donated as $n_1$ and $n_2$($m_1$ and $m_2$), respectively.}
	\label{fig:Cycle}
\end{figure*}

Alice and Bob sacrifice a part of signal pulses as bright reference pulses periodically and send them to Charlie for relative phase calculation. We divide each cycle (2~us) into four regions: a signal region, a recovery region, and two reference regions, as shown in FIG~\ref{fig:Cycle}. Alice and Bob send the bright reference pulses with the intensity of $\mu_r$ to Charlie for interference in the reference region. Alice (Bob) loads 0 (0) and 0 ($\pi/2$) phase in the former and latter reference region, respectively.

In each cycle, Charlie record the counts of SNSPD1 and SNSPD2 in the former and latter reference region, donated as $n_1$, $n_2$, $m_1$ and $m_2$, respectively. Charlie collects counts in time of $T$ around an effective event of X-basis, and the total counts are donated as $N_1$, $N_2$, $M_1$ and $M_2$, respectively.

\begin{equation}
\begin{split}
\label{equ:E3.3}
N_{1,2} = \sum{n_{1,2}},\quad M_{1,2} = \sum{m_{1,2}}
\end{split}
\end{equation}

The relative phase $\delta$ can be calculated by
\begin{align}	
\begin{split}
	&\left\{  
	\begin{aligned}
	& \dfrac{N1}{N1+N2}=\dfrac{1+cos\delta}{2}\\
	& \dfrac{M1}{M1+M2}=\dfrac{1-sin\delta}{2}
	\end{aligned}  
	\right.
\end{split}
\end{align}

There are a few practical aspects to take into consideration in the calculation. On the one hand, we need to send the bright reference pulses to gain enough counts for rapid calculation. However, when the total count rate of two SNSPDs exceeds about 4~MHz, it is hard to increase continuously through increasing $\mu_r$ in our condition. Besides, the bright reference pulses will lead to more noise due to the long fiber channel's nonlinear effect, which is hard to avoid both in the field test and the lab test. On the other hand, the limit count rate of two SNSPDs requires the longer duration time $T$ of each calculation, and it will induce more measuring error in the relative phase calculation and lead to additional QBER in our work. 

It is an optimal trade-off that $\mu_r$ is set to about 450 photons per pulse at a repetition rate of 200~MHz, which results in 5~MHz total count rate of two SNSPDs for calculation. The duration time $T$ of each calculation is 20~us. After filtered by four 100~GHz dense wavelength division multiplexers (DWDMs) in Charlie, the remaining noise is about $1.4 \times 10^{-8}$/pulse.

\subsection{The polarization auto-alignment}
To real-time compensate polarization drifts in the long fiber, we use a polarization auto-alignment module in each node. The schematic of the module is shown in FIG.1 of the main text. 

In Alice (Bob), the polarization mismatch beam is split in a polarizing beam splitter (PBS) and monitored by a commercial power meter. The measurement results are used as the reference for Alice (Bob) to dynamically adjust the DC loaded in an EPC before the PBS. 

In Charlie, slightly different from Alice (Bob), the polarization mismatch part, which is split in two PBSs, are combined with a 70:30 beam splitter (BS) and monitored by a superconducting nanowire single-photon detector SNSPD3. Charlie monitors the reflectivity for the PBSs in real-time during the QKD process. 

Charlie records the average counting rate of SNSPD1 and calculates the polarization alignment efficiency of every minute. We post select the obtained measurement results when the polarization alignment efficiency exceeds 82\%. The probability distribution of the reflectivity for the PBSs in Charlie is shown in FIG. 5. (c) in the main text. The polarization alignment efficiency of all modules is about 94\%.

\subsection{Time Synchronization}
Three 100~KHz electric signals in phase are modulated by two arbitrary-function generators (Tektronix, AFG3253) in Charlie, one of which is used for Charlie's system clock. The rest are used as triggers for two auxiliary synchronization lasers (Sync Lasers) to generate 1570~nm Sync pulses. Multiplexed with the CW bright beam from the master laser, the pulses from each Sync Laser are transmitted through the synchronization channel to Alice and Bob, respectively. Two EDFAs are placed in Yiyuan city (118°12'24.160"E, 36°11'12.602"N) and Zhucheng city (119°24'43.578"E, 36°2'59.305"N). Each EDFA produces a 10~dB gain in both the CW bright beam and the Sync pulses. In Alice's (Bob's) apparatus, the Sync pulses are demultiplexed by a 100~GHz dense wavelength division multiplexer (DWDM) and detected by a photoelectric detector (PD1). The output electric signals are used for regenerating Alice's (Bob's) own 312.5 MHz system clock.

We need to realize indistinguishability in the arriving time of the signal pulses from two independent systems. Before the QKD process, Charlie measures the arriving time of the signal pulses from Alice and Bob, respectively, with a precision of 20~ps. Based on the arrival time's different values, Charlie adjusts the time delay between the electric signals from the arbitrary-function generators. To achieve a better signal to noise ratio, we adopt a time window of 340~ps. 

On account of the long optical fiber length drifting, the arriving time drifts during the QKD process. Charlie monitors the ratio of the non-overlapping between the signal pulse and the detection window in real-time. We post select the obtained measurement results when the ratio of the non-overlapping is less than 55\%. The probability distribution of the ratio is shown in FIG. 5. (d) in the main text. The non-overlapping between signal pulse and detection window in the experiment is 30\%.

\section{DETAILED EXPERIMENTAL RESULTS}

Here we list the experimental results in the following. Table~\ref{tab:para} shows the experimental parameters used by Alice and Bob, and the key length calculation. Table~\ref{tab:totalsendandgain} illustrate the sending and received statistics of all the signals. 

\begin{table*}[htbp]
  \centering
  \caption{Experimental parameters and the key length calculation.}
    \begin{spacing}{2.5}
    \vspace{0.25cm}
    \begin{tabular}{cc}
    \hline
    \hline
    $\mu_{A2}$ & 0.454  \\
    $\mu_{B2}$ & 0.425  \\
    $\mu_{A1}$ & 0.042  \\
    $\mu_{B1}$ & 0.029  \\
    $\epsilon_A$ & 0.307  \\
    $\epsilon_B$ & 0.241  \\
    total number of signal pulse pairs $N$ & 5590517734411  \\
    \hline
    sifted key bits in the Z-basis before AOPP & 27921308  \\
    $QBER_{ZZ}$ - Before AOPP & 27.84\% \\
    \hline
    effective events in X windows & 43382  \\
    QBER in X windows & 9.62\% \\
    \hline
    the relative PLOB bound & 5.01E-09 \\
    the absolute PLOB bound & 1.78E-08 \\
    Key Rate & 4.80E-08 \\
    \hline
    \hline
    \end{tabular}%
  \label{tab:para2}%
  \end{spacing}
\end{table*}%

\begin{table*}[htbp]
  \centering
  \caption{Sending and received statistics of all the signals. Here, the notation $Z_{AO}$ and $Z_A$ ($Z_{BO}$ and $Z_B$) shown in the first column denotes Alice (Bob) chose Z basis and chose the source vacuum and $ u_{A2}$, respectively. The notation $X_{AO}$ and $X_{A1}$ ($X_{BO}$ and $X_{B1}$) denotes Alice (Bob) chose X basis and chose the source vacuum and $ u_{A1}$, respectively.}
      \begin{spacing}{2.5}
      \vspace{0.25cm}
    \begin{tabular}{p{3cm}<{\centering}|p{4cm}<{\centering}p{3cm}<{\centering}}
     \hline
     \hline 
       Source   & Total Send & Total Gain \\
     \hline     
    $Z_{AO}Z_{BO}$ & 1971056824075  & 91307 \\
    $Z_{AO}X_{BO}$ & 53109918477  & 2506 \\
    $Z_{AO}X_{B1}$ & 523112730863  & 622318 \\
    $Z_{AO}Z_B$ & 626936631645  & 10353195 \\
    $X_{AO}Z_{BO}$ & 58301113516  & 2666 \\
    $X_{AO}X_{BO}$ & 1597290781  & 75 \\
    $X_{AO}X_{B1}$ & 15573585117  & 18484 \\
    $X_{AO}Z_B$ & 18768166680  & 309128 \\
    $X_{A1}Z_{BO}$ & 569833486215  & 632696 \\
    $X_{A1}X_{BO}$ & 15174262422  & 17086 \\
    $X_{A1}X_{B1}$ & 151343301524  & 343572 \\
    $X_{A1}Z_B$ & 181292503673  & 3234733 \\
    $Z_{A}Z_{BO}$ & 872120766568  & 9794704 \\
    $Z_{A}X_{BO}$ & 23560039024  & 265688 \\
    $Z_{A}X_{B1}$ & 231207840587  & 2823218 \\
    $Z_{A}Z_B$ & 277529273244  & 7682102 \\
    \hline 
    \hline 
    \end{tabular}%
  \label{tab:totalsendandgain}%
  \end{spacing}
\end{table*}%
%

\end{document}